# ECLIPTIC PROXIMITY AND CLUSTERING OF FAST RADIO BURSTS

SUBHASH N. KARBELKAR

Raman Research Institute, Sir C. V. Raman Avenue, Sadashivanagar, Bengaluru 560080, India

On sabbatical leave from Birla Institute of Technology and Science, Pilani, India

snkarbelkar@gmail.com

Draft version July 23, 2014

## ABSTRACT

It is pointed out that the positional distribution of Fast Radio Bursts (FRBs) shows proximity and clustering around the ecliptic. The FRB's dispersion measure (DM) has a weak (80% significance level) correlation with the magnitude of the ecliptic latitude. The so called Perytons, however, show no significant proximity to any one of the three astronomical planes, viz celestial, ecliptic or galactic, although they are closer to the galactic plane.

## 1. INTRODUCTION

Eight FRBs have been reported so far (Lorimer et al. 2007; Keane et al. 2012; (x4) Thornton et al. 2013; Spitler et al. 2014; Burke-Spolaor et al. 2014) in six surveys. However, no detections were reported in four surveys (Manchester et al. 2001; Burgay et al 2013; Deneva et al. 2009; Burke-Spolaor et al. 2011b). Source models, ranging from terrestrials, galactic to cosmological, have been investigated. The positions of the FRBs can provide a constraint on their models. The positional distribution of the FRBs has been studied by Burke-Spolaor et al. 2014 who consider two exclusive categories for the distribution of FRBs. They club terrestrial, solar system and galactic distributions into the first category of "local/isotropic population". The second category considered by them is "cosmological population". Since different radio burst surveys have different parameters, such as the survey hours, sky coverage, DM search limit etc they work with the surveys as the basic units in their analysis. In order to distinguish between the two categories, considered by them, they have developed a model to predict the number of events each survey should have detected. The present analysis differs from theirs in two main aspects. Firstly, we do not assume that a solar system origin for FRBs must lead to a uniform sky distribution. Kulkarni et al. 2014, for example, have discussed a possibility that Jupiter may be the source behind the chirped signal. Although dismissed by them on the grounds of remoteness of the planet at the time of the burst, such a possibility will tie the positions of the FRBs to the orbit of the planet leading to a high correlation in the positions, in particular to a proximity to the ecliptic plane within beam size. A second order process, such as reflection off an intermediate object, such as an aero plane considered by Kulkarni et al. 2014, will lead to spread in the directions. However, as the laws of reflections have to be obeyed, the resulting distributions of reflections may not become uniform; some correlation may still survive. Even if the reflection is from a rough round surface the observed flux will depend on how much of the surface is illuminated and how much of that is visible at the location of the observer. Until the nature of the FRB phenomena is known one cannot rule out positional non-uniformity, in case the phenomenon owes its origin, in parts, to the solar system.

Our second point of departure, from the analysis by Burke-Spolaor et al. (2014), is that we do not estimate how many FRBs should one have detected but deal with the known positions of the FRBs. For this purpose only those beams, in any survey, which contain detection count. A survey dependent parameter that could have been relevant is the beam size. However, beam sizes in all the surveys turn out to be negligible compared to the positional statistic (mean and standard deviation) involved in this analysis.

## 2. OUR STRATEGIE

Burke-Spolaor et al. (2014) have partly excluded the detection by Keane et al. (2012) on the ground that a report by Bannister & Madsen (2014) shows it to be a galactic with high probability. Since we are open to a possibility in the solar system we include this detection. As far as the detection by Lorimer et al. (2007) is concerned, we subscribe to the ambiguity, whether an FRB or a Peryton (Burke-Spolaor et al. 2011a), in its nature and follow Burke-Spolaor et al. (2014) to exclude it from consideration.

Our strategy in getting a clue to the possible origin of the FRBs (or perytons) is to ask the question which, if any, is the "natural" coordinate system for the FRBs. The equator of the earth, which define the celestial plane, is in special position for phenomena involving Earth's rotation. The solar system lies essentially in the ecliptic plane. The Milky way defines the galactic plane. These planes are inclined relative to each other and it should be possible to count on events where the planes are well separated to single out the "natural" plane and, therefore, the "natural" origin/significant factor. If the FRB phenomenon has to do with any one of these three factors we expect some clustering around the corresponding plane (or the pole). To this end we find the coordinate system in which they are closest to the equatorial plane. As another indicator for the "natural" coordinate system we also find the coordinate system, among these three, in which they all together occupy (as defined by the four extreme coordinate values) the smallest area. We also carry out similar analysis for the Perytons (Burke-Spolaor et. al 2011a).

## 3. FRB POSITIONS AND DM CORRELATION

In Table 1 we list the Latitude and Longitude coordinates, in the three coordinate systems relevant to astronomy; namely the celestial (based on the spin of the earth), the ecliptic (based on the plane of the earth's orbit (approximately the plane of the solar system objects)) and the galactic (based on the plane of the galaxy) coordinate system, of the seven FRB events considered. Since longitude is cyclic in some places 360 has been either added (marked by an asterisk) or subtracted (marked by a hash) to keep all numbers in the same cycle. The longitudes are listed for completeness only, as only the latitude is relevant in deciding the natural plane. We see from the table 1 that it is the Ecliptic coordinate system in which the mean Latitude has the smallest value of -5.1$^o$. For the celestial and the galactic systems the mean Latitude is, significantly larger, -13.5$^o$ and -35.1$^o$ respectively. It can be seen from the table that if a FRB is close to either of the celestial or galactic planes, there are three such instances, then it is also within 15$^o$ of the ecliptic. The last row of the Table 1 gives the expected standard deviation in the corresponding mean values. In estimating these values the standard deviation listed in the previous row is treated as an estimator for the population value of the standard deviation in the latitude. The mean for 7 values, therefore, can fluctuate by the population standard deviation divided by $\sqrt{6}$ (one degree of freedom removed on account of mean also being estimated from the same data). We infer that the mean Latitude -5$^o$ in the Ecliptic coordinate, given the expected standard deviation of 6, is not significantly different from 0. For the other two, the celestial and the galactic, coordinate systems the mean deviates from zero (planar value) respectively by 1.3 times and 3.2 times the corresponding expected standard deviation in the mean.

From the Table 1 we also see that not only in the mean sense but for as many as six of the FRBs (FRB010621 being the sole exception) the ecliptic latitude is less than (significantly so in some cases) that in the other two coordinate systems. This observation makes our inference robust against a possible variation in relative weights that can be assigned, as a means to compensate a possible selection bias, to different FRBs. Although one may give different weights to different FRBs, for a given FRB, the weight for its three latitude values (in 3 systems) will be the same and since the ecliptic value is smaller than others for 6 of them the mean will also be so.

We also note that the dispersion measure (DM) has a weak correlation with the magnitude of ecliptic latitude. This is seen clearly in the case of FRB110703 which is closest to the ecliptic (0.3$^o$) and has the largest DM 1073 pc cm$^{-3}$; and in the case of the next closest (-3$^o$) FRB110220 which has the second largest DM of 910 pc cm$^{-3}$. The corresponding correlation coefficient, -0.55, yields a value 1.475 for the (Student's) t value. This implies a probability of 20% for the null hypothesis, that there is no linear relation between the magnitude of the ecliptic latitude and the DM, to be true.

Table 1

The Julian Date, DM, Latitude and longitude coordinates of the FRBs in the three coordinate systems.

| FRB….. | JD 245…. | DM | Latitude | | | Longitude | | |
|---|---|---|---|---|---|---|---|---|
| | | | celestial | ecliptic | galactic | celestial | ecliptic | galactic |
| 010621 | 2082 | 746 | -8.5 | 14.6 | -4 | 283 | 283.3 | 25.4 |
| 110220 | 5613 | 910 | -12.4 | -3 | -54.6 | 338 | 335.3 | 50.7 |
| 110627 | 5740 | 677 | -44.7 | -26.7 | -41.6 | 315 | 304.7 | -4.1[#] |
| 110703 | 5746 | 1072 | -2.9 | 0.3 | -58.9 | 352.5 | 352 | 80.6 |
| 110703 | 5954 | 521 | -18.4 | -12.5 | -66.2 | 348.8 | 342.4 | 49.3 |
| 120127 | 6234 | 557 | 33 | 9.8 | -0.3 | 443[*] | 444.1[*] | 174.9 |
| 011025 | 1915 | 790 | -40.6 | -18 | -20 | 286.8 | 283.3 | -3.4[#] |
| | | | | | | | | |
| Mean | | | -13.5 | -5.1 | -35.1 | 338.2 | 335.0 | 53.3 |
| Std Dev | | | 25.9 | 14.9 | 27.0 | 54.0 | 55.5 | 61.7 |
| Std dev in mean | | | 10.6 | 6.1 | 11.0 | | | |

Table 2

Four corners of a "cover" (minimal sector of coordinate system) which encloses all the seven FRBs considered

| | Latitude range | | Longitude Range | | Solid angle |
|---|---|---|---|---|---|
| | Highest | Lowest | From | To | |
| Celestial | 33 | -44.7 | 283 | 443[*] | 3.48 |
| Ecliptic | 14.6 | -26.7 | 283.3 | 444.1[*] | 1.97 |
| Galactic | -0.3 | -66.2 | -3.4[#] | 174.9 | 2.83 |

We see from Table 2 that it is in the Ecliptic coordinate system that the area occupied by the distribution of the FRBs has the smallest "cover" as defined by the four extremes in the corresponding coordinate system. This is yet another indication that the ecliptic system is, perhaps, a more natural system for FRBs.

The positional distribution in longitude, in any coordinate system is far from uniform. This is expected as surveys in the region of galactic plane don't cover the entire longitude range. We note that in the case of solar system mediated occurrence the longitude symmetry is expected to manifest itself only on time scales larger than those characteristic of the system.

## 4. PERYTON POSITIONS

In table 3 we present positional coordinates of the so called Perytons in the three natural astronomical coordinate systems. In contrast to the case of FRBs there is no significant "natural" association with any of the three coordinate systems. The mean latitude is smallest in the galactic coordinate system, however, this mean is larger than the standard deviation in the mean one expects on the basis of the observed standard deviation in latitude of the Perytons. This conclusion holds whether we treat the first listed event as a single long duration intermittent event or a set of ten events. The observed separation between the positions of the FRBs and the Peryton should pose a challenge to any unified theory of the two phenomena. However, Burke-Spolaor et al. 2011a have reported that the Perytons detections were made through a side lobe $>>5^o$ away from the pointing direction. There is some possibility, here, to reconcile the discrepancy in the locations of the two types of burst if the Perytons originated, like the FRBs, in the vicinity of the ecliptic but were detected via a far away side lobe.

Table 3

Latitude and longitude coordinates of the Peryton events in the three coordinate systems. The first event is actually a decuple event spanning about 245 seconds.

|  | Latitude | | | Longitude | | |
|---|---|---|---|---|---|---|
|  | celestial | ecliptic | galactic | celestial | ecliptic | galactic |
| Decuple event | -52.2 | -66.5 | -7.3 | 128.2 | 161.4 | 269 |
|  | -50.8 | -64.6 | -5.4 | 130.2 | 161.8 | 268.6 |
|  | -50.7 | -50 | 8.9 | 167 | 196 | 286.7 |
|  | -4.5 | 14.3 | -23 | -53.5 | -52.3 | 399.9 |
|  | -60.9 | -84.3 | -29.7 | 89.9 | 89.5 | 269.8 |
|  | 10.8 | -11.3 | -22.4 | 70.6 | 70.5 | 186.9 |
|  | Statistics counting the first event only once | | | | | |
| Mean | -34.7167 | -43.7333 | -13.15 | 88.73333 | 104.4833 | 280.15 |
| Std Dev | 29.96728 | 37.57247 | 14.40122 | 77.41224 | 90.39954 | 68.4739 |
| Std dev in Mean | 13.4 | 16.8 | 6.4 |  |  |  |

|  | Statistics counting the first event ten times | | | | | |
|---|---|---|---|---|---|---|
| Mean | -45.2 | -57.4 | -9.6 | 112.4 | 138.6 | 273.5 |
| Std Dev | 20.0 | 25.2 | 9.1 | 50.4 | 61.2 | 41.3 |

## 5. SUMMARY

We conclude that the distribution of the seven (out of 8 known so far) FRBs shows significant proximity and clustering around the ecliptic. This result can be interpreted as indicating a role, more significant than credited so far, the solar system plays in their occurrence. We also find a weak correlation between their DM and their ecliptic latitude; another hint at the role solar system may lay in the phenomena. Perytons, on the other hand show no such "natural" coordinate system though they are closer to the galactic plane. This positional separation, between the two types of bursts, poses a challenge to a unified theory of the FRB and Peryton phenomena unless one admits a possibility that the Perytons are FRBs detected via side lobe/s very far the from pointing direction. We note that these conclusions are based on the observed positions of FRBs and Perytons and detailed statistical modeling is needed to take into account effects of sky coverage in different surveys. Till such analysis is done these conclusions serve as a pointer to a possible role of the solar system in these radio bursts.

## ACKNOWLEDGEMENTS

It is a great pleasure to acknowledge S. Burke-Spolaor for providing the RAs and DECs for the Perytons.